\documentclass[preprint,aps,prc,showpacs,nofootinbib]{revtex4}%
\usepackage{amssymb}
\usepackage{epsfig}
\usepackage{amsmath,graphicx}
\usepackage{axodraw}
\usepackage{amsmath}
\usepackage{amsfonts}
\usepackage{graphicx}%
\providecommand{\U}[1]{\protect\rule{.1in}{.1in}}

\newcommand\ba{\begin{eqnarray}}
\newcommand\ea{\end{eqnarray}}

\begin{document}
\title{QED radiative corrections to the decay $\pi^{0}\rightarrow e^{+}e^{-} $ }
\author{A. E. Dorokhov, E. A. Kuraev, Yu. M. Bystritskiy and M. Se\v cansk\'y}
\affiliation{\textit{JINR-BLTP, 141980 Dubna, Moscow region, Russian Federation}}
\date{\today }

\begin{abstract}
We reconsider QED radiative corrections (RC) to the $\pi^{0}\rightarrow
e^{+}e^{-}$ decay width. One kind of RC investigated earlier has a
renormalization group origin and can be associated with the final state
interaction of electron and positron. It determines the distribution of lepton
pair invariant masses in the whole kinematic region. The other type of RC has
a double-logarithmic character and is related to almost on-mass-shell behavior
of the lepton form factors. The total effect of RC for the $\pi^{0}\rightarrow
e^{+}e^{-}$ decay is estimated to be $3.2\%$ and for the decay $\eta
\rightarrow e^{+}e^{-}$ is $4.3\%$.

\end{abstract}
\maketitle


\section{Introduction}

Rare decays of mesons serve as the low-energy test of the Standard Model.
Accuracy of experiments has increased significantly in recent years.
Theoretically, the main limitation comes from the large distance contributions
of the strong sector of the Standard Model where the perturbative theory does
not work. However, in some important cases the result can be essentially
improved by relating these poorly known contributions to other experimentally
known processes. The famous example is the Standard Model calculation of the
anomalous magnetic moment of muon $\left(  g-2\right)  _{\mu}$ where the data
of the processes $e^{+}e^{-}\rightarrow hadrons$ and $\tau\rightarrow hadrons$
are essential to reduce the uncertainty. It turns out that this is also the
case for the rare neutral pion decay into an electron-positron pair measured
recently by the KTeV collaboration \cite{Abouzaid07} and reconsidered
theoretically in \cite{DI07}.

The measured branching is \cite{Abouzaid07}%
\begin{equation}
B^{\mathrm{KTeV}}\left(  \pi^{0}\rightarrow e^{+}e^{-},x_{D}>0.95\right)
=\left(  6.44\pm0.25\pm0.22\right)  \cdot10^{-8},\label{BktevX}%
\end{equation}
where the kinematic cut over the Dalitz variable $x_{D}\equiv\left(
p_{+}+p_{-}\right)  ^{2}/M^{2},$ $\nu^{2}\equiv4m^{2}/M^{2}\leq x_{D}\leq1$,
was used in order to suppress the Dalitz decay events $\pi^{0}\rightarrow
e^{+}e^{-}\gamma$. Then, the important step in extraction of the branching
consists in correct treating the radiative corrections (RC) to the process
which has been considered earlier in \cite{B83} and \cite{T93}. Extrapolating
the full radiative tail beyond $x_{D}>0.95$ and scaling the result back up by
the overall RC leads to the final result
\cite{Abouzaid07}
\begin{equation}
B_{0}^{\mathrm{KTeV}}(\pi_{0}\rightarrow e^{+}e^{-})=\left(  7.49\pm
0.29\pm0.25\right)  \cdot10^{-8},\label{Bktev}%
\end{equation}
where the leading order radiative corrections have been taken into account
\cite{B83}. It is the motivation of our paper to revise the calculation of QED
RC to the $\pi_{0}\rightarrow e^{+}e^{-}$ decay width.

\begin{figure}[th]
\includegraphics[width=10cm]{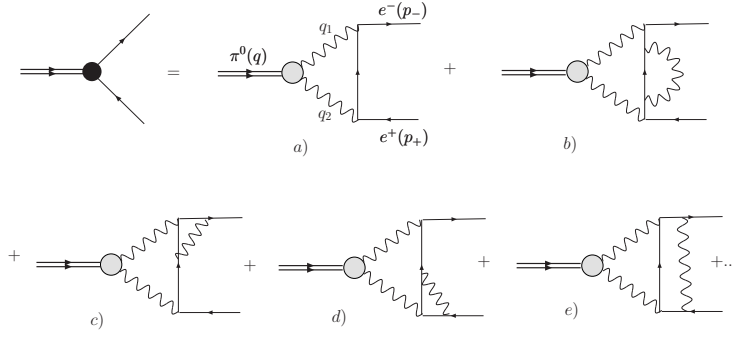}\caption{Set of the lowest order QED
RC to $\pi^{0}\rightarrow e^{+}e^{-}$ process: virtual corrections.}%
\label{fig:triangle}%
\end{figure}

In the lowest order of QED perturbation theory (PT), the photonless decay of
the neutral pion,
\[
\pi_{0}(q)\rightarrow e^{-}(p_{-})+e^{+}(p_{+}),\quad q^{2}=M^{2},\quad
p_{\pm}^{2}=m^{2},
\]
($M$ meson mass, $m$ lepton mass) is described by the one-loop Feynman
amplitude (Fig. 1a) corresponding to the conversion of the pion through two
virtual photons into an electron-positron pair. The normalized branching ratio
is given by \cite{Dr59,BG60,B82}
\begin{equation}
R_{0}(\pi_{0}\rightarrow e^{+}e^{-})=\frac{B_{0}\left(  \pi^{0}\rightarrow
e^{+}e^{-}\right)  }{B\left(  \pi^{0}\rightarrow\gamma\gamma\right)  }%
=2\beta\left(  \frac{\alpha m}{\pi M}\right)  ^{2}|\mathcal{A}\left(
M^{2}\right)  |^{2},\label{Bpi}%
\end{equation}
where $\beta=\sqrt{1-\nu^{2}},$ $B\left(  \pi^{0}\rightarrow\gamma
\gamma\right)  =0.988$ and the reduced amplitude is
\begin{equation}
\mathcal{A}\left(  q^{2}\right)  =\frac{2}{M^{2}}\int\frac{d^{4}k}{i\pi^{2}%
}\frac{(qk)^{2}-q^{2}k^{2}}{(k^{2}+i\epsilon)\left[  (q-k)^{2}+i\epsilon
\right]  \left[  (p_{-}-k)^{2}-m^{2}+i\epsilon\right]  }F_{\pi}(-k^{2}%
,-(q-k)^{2}),\label{Rq}%
\end{equation}
with the pion transition form factor $F_{\pi}(-k^{2},-q^{2})$ being normalized
as $F_{\pi}(0,0)=1$. The imaginary part of $\mathcal{A}\left(  q^{2}\right)  $
can be found in a model independent way \cite{BG60}%
\begin{equation}
\mathrm{Im}\mathcal{A}(M^{2})=-\frac{\pi}{2\beta}\ln\left(  \frac{1+\beta
}{1-\beta}\right)  ,\label{ImR}%
\end{equation}
while the real part is reconstructed by using the dispersion approach up to a
subtraction constant%
\begin{equation}
\mathrm{Re}\mathcal{A}(M^{2})=\mathcal{A}(0)+\frac{1}{\beta}\left[  \frac
{\pi^{2}}{12}+\frac{1}{4}\ln^{2}\left(  \frac{1+\beta}{1-\beta}\right)
\right]  .\label{ReR}%
\end{equation}

Usually this constant, containing the nontrivial dynamics of the process, is
calculated within different models describing the form factor $F_{\pi}%
(k^{2},q^{2})$ \cite{B82,Ba83,SL92,DI07}. However, it has recently been shown
in \cite{DI07} that this constant may be expressed in terms of the inverse
moment of the pion transition form factor given in symmetric kinematics of
spacelike photons%
\begin{equation}
\mathcal{A}\left(  q^{2}=0\right)  =3\ln\left(  \frac{m_{e}}{\mu}\right)
-\frac{3}{2}\left[  \int_{0}^{\mu^{2}}dt\frac{F_{\pi\gamma^{\ast}\gamma^{\ast
}}\left(  t,t\right)  -1}{t}+\int_{\mu^{2}}^{\infty}dt\frac{F_{\pi\gamma
^{\ast}\gamma^{\ast}}\left(  t,t\right)  }{t}\right]  -\frac{5}{4}.\label{R0}%
\end{equation}
Here, $\mu$ is an arbitrary (factorization) scale. One has to note that the
logarithmic dependence of the first term on $\mu$ is compensated by the scale
dependence of the integrals in the brackets. The accuracy of these
calculations are determined by omitted small power corrections of the order $O(\frac
{m^{2}}{\Lambda^{2}})$ and $O(\frac{m^{2}}{M^{2}}L)$ in the r.h.s.
(\ref{ReR}), where $\Lambda\lesssim M_{\rho}$ is the characteristic scale of
the form factor $F_{\pi\gamma^{\ast}\gamma^{\ast}}\left(  t,t\right)  $ and
$L$ is the large logarithm parameter%
\[
L=\ln\left(  \frac{M^{2}}{m^{2}}\right)  \approx\ln\left(  \frac{1+\beta
}{1-\beta}\right)  .
\]
For the decay $\pi^{0}\rightarrow e^{+}e^{-}$ one has $L\approx11.2$.

By using the representation (\ref{R0}), and the CELLO \cite{Behrend:1990sr}
and CLEO \cite{Gronberg:1997fj} data on the pion transition form factor
$F_{\pi\gamma^{\ast}\gamma^{\ast}}^{\mathrm{CLEO}}\left(  t,0\right)  $ given
in asymmetric kinematics the lower bound on the decay branching ratio was
found in \cite{DI07}. This lower bound follows from the property:
$F_{\pi\gamma^{\ast}\gamma^{\ast}}\left(  t,t\right)  <F_{\pi\gamma^{\ast
}\gamma^{\ast}}\left(  t,0\right)  $ for $t>0$. It considerably improves the
so-called unitary bound obtained from the property $\left\vert \mathcal{A}%
\right\vert ^{2}\geq\left(  \operatorname{Im}\mathcal{A}\right)  ^{2}$.
Further restrictions follow from QCD and allow one to make a model independent
prediction for the branching  \cite{DI07}
\begin{equation}
B_{0}^{\mathrm{Theor}}\left(  \pi^{0}\rightarrow e^{+}e^{-}\right)  =\left(
6.2\pm0.1\right)  \cdot10^{-8},\label{Bth}%
\end{equation}
which is $3.3\sigma$ below the KTeV result (\ref{Bktev}). The main source of
the error in (\ref{Bth}) is defined by indefiniteness in the knowledge of the
pion form factor $F_{\pi\gamma^{\ast}\gamma^{\ast}}\left(  t,t\right)  $  \cite{DI07}. The
discrepancy between (\ref{Bth}) and (\ref{Bktev}) requires further attention
from experiment and theory to this process because there are not many places
where experiment is in conflict with the Standard Model.

Considering the higher orders of QED\ PT, there are two sources of RC to the
width of the $\pi^{0}\rightarrow e^{+}e^{-}$ decay (Fig. 1). One of them has a
renormalization group origin and can be associated with the final state
interaction of electron and positron. The relevant contribution corresponds to
taking into account the charged particle interaction at large distances. The
other is of double-logarithmic character and is related to short distance
contributions. Let us note that the branching ratio (\ref{Bpi}) is
proportional to the electron mass squared and thus the
Kinoshita--Lee-Nauenberg theorem of cancellation of mass singularities in the
limit $m\rightarrow0$ is not violated.

\begin{figure}[th]
\includegraphics[width=10cm]{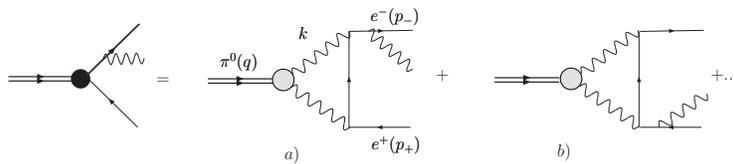}\caption{RC due to soft photon
emission.}%
\end{figure}

The first kind of corrections was considered in \cite{B83}. Later, the effect
of the higher order RC was estimated in \cite{T93} by using the exponentiation
of soft photon contributions, which is essentially equivalent to the Yennie,
Frautchi-Suura factorization procedure. Considering only the
two-virtual-photon conversion to a lepton pair, it is originated from the box
type Feynman amplitude (Fig. 1e) as well as the contribution from the emission
of real photons by leptons (Fig. 2) and produces the single-logarithmic
enhanced terms ($\sim L$) which are described by the lepton nonsinglet
structure function method \cite{KF85}. It was shown in \cite{T93} that the
soft photon emission can drastically change the results obtained at the Born
level when the invariant mass of leptons is close to the pion mass.

Furthermore, we find an additional source of RC which is of the so-called
"double-logarithmic" (DL) nature $\left(  \alpha L^{2}/\pi\sim1\right)  $.
This kind of asymptotics was intensively investigated in the 70s in a series
of QED\ processes \cite{AB81,BLP92}. The DL type contribution to the decay
width was not considered earlier for the $\pi^{0}\rightarrow e^{+}e^{-}$ decay.

\section{Large logarithm regime and double-logarithmic correction}

First, it is instructive to reproduce the results discussed above in a simple
and physical way. For this aim we note that the main contribution to the real
part of $\mathcal{A}\left(  M^{2}\right)  $ comes from the kinematic region of
loop momenta corresponding to the intermediate virtual electron (or positron)
close to the mass shell (Fig. 1a). Really, by changing the integration
variable in (1) as $q_{1}=k=p_{-}-\kappa,\quad q_{2}=q-k=p_{+}+\kappa$ and
omitting terms of order $O(\kappa^{2}/M^{2})$ we can rewrite the amplitude in
the Born approximation $\mathcal{A}_{0}\left(  M^{2}\right)  $ as
\[
\mathcal{A}_{0}\left(  M^{2}\right)  \approx\frac{1}{2}\int\frac{d^{4}\kappa
}{i\pi^{2}}\frac{M^{2}}{(\kappa^{2}-m^{2}+i\epsilon)((\kappa-p_{-}%
)^{2}+i\epsilon)((\kappa+p_{+})^{2}+i\epsilon)}.
\]
Let us find the real part of the amplitude within the leading logarithmic
accuracy that corresponds to the restriction of the kinematic region by
conditions
\begin{equation}
m^{2}\approx|\kappa|^{2}\ll\left(  |q_{1}^{2}|,|q_{2}^{2}|\right)  \ll
M^{2}.\label{Status}%
\end{equation}
To this end one performs the substitutions
\begin{align}
&  \frac{d^{4}\kappa}{\Delta+i\epsilon}\rightarrow\frac{-i\pi\overrightarrow
{\kappa}^{2}d|\overrightarrow{\kappa}|dO_{\kappa}}{2\omega}\left.  \left[
\Theta\left(  \kappa_{0}\right)  +\Theta\left(  -\kappa_{0}\right)  \right]
\right\vert _{\kappa_{0}^{2}=\omega^{2}},\quad\label{MassShellApproximation}\\
&  q_{1}^{2}=-2M\kappa_{0}u,\quad q_{2}^{2}=2M\kappa_{0}\left(  u+\beta
_{\omega}\cos\theta\right)  ,\nonumber\\
&  \omega=\sqrt{\overrightarrow{\kappa}^{2}+m^{2}},\quad\Delta=\kappa
^{2}-m^{2},\quad\beta_{\omega}=\sqrt{1-\frac{m^{2}}{\omega^{2}}},\qquad
u=\frac{1-\beta_{\omega}\cos\theta}{2}\nonumber
\end{align}
where $\theta$ is the angle between the directions of electron momentum (the
rest frame of the initial pion implied) and the 3-momentum of the virtual
electron. Let us note that by kinematical reasons in the region of maximal
contribution the signs of  $q_{1}^{2}$ and $q_{2}^{2}$ must be opposite.
Performing the angular integration we obtain the leading term of (\ref{ReR})%

\begin{equation}
\mathrm{\operatorname{Re}}\mathcal{A}_{0}\left(  M^{2}\right)  =-\frac{M^{2}%
}{\pi}\int\frac{d^{3}\kappa}{\omega q_{1}^{2}q_{2}^{2}}=\int_{m}^{\frac{M}{2}%
}\frac{\beta_{\omega}d\omega}{\omega}\ln\frac{\omega^{2}}{m^{2}}\approx
\frac{1}{4}L^{2}. \label{ReA0}%
\end{equation}

Then, let us consider the vertex type RC (Fig. 1 c,d). The lowest order
evaluation arising from the diagrams of Figs. 1b) and 1c) leads to the
correction $\Gamma(q_{1},\kappa,p_{-})=1-\frac{\alpha}{2\pi}I_{V}(q_{1}%
^{2},\kappa^{2})$ with
\[
I_{V}(q_{1}^{2},\kappa^{2})=\ln\frac{|q_{1}^{2}|}{m^{2}}\ln\frac{|q_{1}^{2}%
|}{\left\vert {\kappa^{2}}\right\vert }+\frac{1}{2}\ln^{2}\frac{|q_{1}^{2}%
|}{\left\vert \kappa^{2}\right\vert }-\frac{3}{2}\ln\frac{|q_{1}^{2}|}{m^{2}%
}+\frac{\pi^{2}}{3}+\frac{1}{2}-\ln\frac{m}{\lambda}-\Theta\left(  -q_{1}%
^{2}\right)  \frac{3\pi^{2}}{2},
\]
which is consistent with the result of similar calculations in \cite{NLL}. The
last term arises from a renormalization procedure. A similar contribution
$I_{V}(q_{2}^{2},\kappa^{2})$ comes from the diagrams of Figs. 1b) and 1d).

Let us consider now the box-type diagram (see Fig. 1e) and demonstrate the
calculations in more detail. The corresponding contribution to the amplitude
has the form
\begin{equation}
\frac{\Delta}{2}\operatorname{Re}\int\frac{d^{4}k_{1}}{i\pi^{2}}\frac
{N}{\left(  k_{1}^{2}-\lambda^{2}+i\epsilon\right)  \left(  (p_{-}+k_{1}%
)^{2}-m^{2}+i\epsilon\right)  \left(  (p_{+}-k_{1})^{2}-m^{2}+i\epsilon
\right)  \left(  (\kappa+k_{1})^{2}-m^{2}+i\epsilon\right)  },\label{IB1}%
\end{equation}
where%
\[
N=\bar{u}(p_{-})\gamma_{\mu}(p_{-}+k_{1}+m)\gamma_{\lambda}(\kappa
+k_{1}+m)\gamma_{\gamma}(-p_{+}+k_{1}+m)\gamma_{\mu}v(p_{+}).
\]
In the leading kinematic region, where $k_{1}$ and $\kappa$ are small, we can
reduce the numerator to
\[
N\approx-2mM^{2}\bar{u}(p_{-})\gamma_{\lambda}\gamma_{\gamma}v(p_{+}).
\]
The calculation of the scalar 4-denominator integral is standard: by using the
Feynman parametrization and performing loop momentum integration we arrive at
\[
-2mM^{2}\Delta\int_{0}^{1}dx\int_{0}^{1}ydy\int_{0}^{1}\frac{z^{2}dz}{\left(
Az^{2}+Bz+C\right)  ^{2}},
\]
with
\begin{align}
A &  =(yp_{x}-\bar{y}\kappa)^{2};\ B=-\bar{y}\Delta-\lambda^{2};\ C=\lambda
^{2},\ \\
p_{x} &  =xp_{+}-\bar{x}p_{-},\quad p_{x}^{2}=m^{2}-M^{2}x\bar{x},\quad\bar
{x}=1-x,\ \bar{y}=1-y.\quad\nonumber
\end{align}
First, we perform the integration in $z$
\begin{align}
\int_{0}^{1}\frac{z^{2}dz}{(Az^{2}+Bz+C)^{2}} &  =\frac{2C+B}{(A+B+C)R}%
-\frac{2C}{R^{\frac{3}{2}}}\ln\frac{2C+B+\sqrt{R}}{2C+B-\sqrt{R}},\label{11}\\
R &  =B^{2}-4AC>0.\nonumber
\end{align}
Calculating the $y$-integral of (\ref{11}) results in
\[
-\frac{1}{p_{x}^{2}}\left[  \frac{1}{2}\ln\frac{\Delta^{2}}{\lambda^{2}m^{2}%
}+\frac{1}{2}\ln\frac{p_{x}^{2}}{m^{2}}-\ln\left\vert \frac{p_{x}^{2}}%
{q_{1}^{2}\bar{x}+q_{2}^{2}x}\right\vert \right]  .
\]
Integration in $x$ by using
\[
\operatorname{Re}\int_{0}^{1}\frac{dx}{p_{x}^{2}+i\epsilon}=-\frac{2}{M^{2}%
}L,\quad\operatorname{Re}\int_{0}^{1}\frac{dx}{p_{x}^{2}+i\epsilon}\ln
\frac{p_{x}^{2}+i\epsilon}{m^{2}}=-\frac{1}{M^{2}}(L^{2}-\frac{4}{3}\pi^{2}),
\]
leads to the correction $\frac{\alpha}{2\pi}I_{B}$ with
\begin{equation}
I_{B}\left(  q_{1}^{2},q_{2}^{2}\right)  =-\frac{1}{2}L^{2}-2(L-1)\ln\frac
{m}{\lambda}-L(L_{1}+L_{2})-\frac{1}{2}(L_{1}-L_{2})^{2}+\frac{1}{2}\pi
^{2},\qquad\label{IBa}%
\end{equation}
where
\[
L_{1,2}=\ln\frac{|q_{1,2}^{2}|}{m^{2}}.
\]

Finally, one needs to integrate over photon momenta $q_{1}$ and $q_{2}$.
Again, the logarithmically enhanced contribution comes from the kinematic
regions $m\leq\omega\leq M/2,$ $\cos\theta\rightarrow\pm1$ (see definitions in
(\ref{MassShellApproximation})). The Born amplitude (one-loop) and the
lowest-order radiative correction to it can be written as (we take into
account the equal contributions of regions $|q_{1}^{2}|\ll|q_{2}^{2}|$ and
$|q_{2}^{2}|\ll|q_{1}^{2}|$)
\begin{equation}
\sim\int_{m}^{\frac{M}{2}}\frac{d\omega}{\omega}\beta_{\omega}\int
_{\frac{m^{2}}{4\omega^{2}}}^{1}\frac{du}{u}\left[  1+\frac{\alpha}{2\pi
}(I_{V}\left(  q_{1}^{2},m^{2}\right)  +I_{V}\left(  q_{2}^{2},m^{2}\right)
+I_{B}\left(  q_{1}^{2},q_{2}^{2}\right)  )\right]  \label{IVB}%
\end{equation}
with (we put here $\Delta\approx m^{2}$)
\begin{align}
I_{V}\left(  q_{1}^{2},m^{2}\right)  +I_{V}\left(  q_{2}^{2},m^{2}\right)
+I_{B}\left(  q_{1}^{2},q_{2}^{2}\right)   &  =-\frac{1}{4}L^{2}-\frac{1}%
{2}Ll-\frac{3}{4}l^{2}-2(L-1)\ln\frac{m}{\lambda}+\frac{3}{2}L+\frac{3}%
{2}l\nonumber\\
&  -l_{u}^{2}-\left(  \frac{1}{2}L+\frac{3}{2}\left(  l-1\right)  \right)
l_{u}+\frac{\pi^{2}}{6}+1,
\end{align}
where we use substitutions (\ref{MassShellApproximation}) and introduce the
notation%
\begin{equation}
l=\ln\left(  \frac{4\omega^{2}}{m^{2}}\right)  ,\qquad l_{u}=\ln\left(
\frac{1-\beta_{\omega}\cos\theta}{2}\right)  .
\end{equation}
Integration of (\ref{IVB}) leads to (we keep only terms of order $L^{2},L$ and
$L^{0}$)
\begin{align}
\frac{R_{virt}}{R_{0}} &  =1+\delta_{virt},\nonumber\\
\delta_{virt} &  =\frac{\alpha}{\pi}\left[  -\frac{13}{24}L^{2}-2(L-1)\ln
\frac{m}{\lambda}+\frac{3}{4}L+\frac{\pi^{2}}{6}+2\right]  .\label{dV}%
\end{align}

Consider now the real photon emission corrections. One can distinguish two
mechanisms of the radiative decay $\pi^{0}\rightarrow e^{+}e^{-}\gamma$. One
of them, the so-called Dalitz process, corresponds to decay mode of the pion
to real and virtual photons with a subsequent decay of the virtual photon to
the $e^{+}e^{-}$ pair. The corresponding contribution to the width is not
suppressed by lepton mass and provides an important background to the $\pi
_{0}\rightarrow e^{+}e^{-}$ process \cite{MS72,B83,KKN06}. However, the Dalitz
matrix element squared and its interference with the double virtual photon
amplitude are suppressed by $\sim\left(  1-x_{D}\right)  ^{3}$ and
$\sim\left(  1-x_{D}\right)  ^{2}$ as $x_{D}\rightarrow1$ \cite{B83} (Fig. 3).
This results in a negligible (of order $0.02\%$) interference contribution
integrated in the region of interest for this measurement, $0.9<x_{D}<1$
\cite{thesises}.

Another mechanism consists in creation of a lepton pair by two virtual photons
with emission of real photon by a pair components. For emission of a soft
photon (with energy $\omega$ not exceeding $\Delta\epsilon<<\frac{M}{2}$ in
the pion rest frame) the standard calculations \cite{AB81} give
\begin{equation}
\delta_{\mathrm{soft}}=\frac{\alpha}{\pi}\left[  2\left(  L-1\right)  \ln
\frac{2\Delta\varepsilon}{M}+2\left(  L-1\right)  \ln\frac{m}{\lambda}%
+\frac{1}{2}L^{2}-\frac{\pi^{2}}{3}\right]  . \label{ds}%
\end{equation}
Emission of a hard photon was investigated in \cite{B83} with the result
\begin{equation}
\delta_{\mathrm{hard}}=\frac{\alpha}{\pi}\left[  -2\left(  L-1\right)
\ln\frac{2\Delta\varepsilon}{M}-\frac{3}{2}\left(  L-1\right)  -\frac{\pi^{2}%
}{3}+\frac{7}{4}\right]  . \label{dh}%
\end{equation}

In order to find the distribution over a lepton pair invariant mass, the
adequate way is to use the method of structure functions \cite{KF85} based on
the application of the renormalization group approach to QED. Here, the
nonsinglet structure function is associated with a final state fermion line.
In partonic language it describes the probability for a fermion to stay a
fermion. Omitting the events with creation of more than one lepton, the
nonsinglet structure function of electron relevant to the process valid at all
orders in perturbation theory is\footnote{Note that the value of the
$K$-factor in (\ref{F2}), $K=1+\frac{3}{4}\frac{\alpha}{\pi},$ differs from
that obtained in \cite{KF85} for the process of hadron production in
single-photon $e^{+}e^{-}$ annihilation channel.}
\begin{equation}
F(x_{D})=b\left(  1-x_{D}\right)  ^{b-1}\left(  1+\frac{3}{4}b\right)
-\frac{1}{2}b(1+x_{D})+O\left(  b^{2}\right)  ,\label{F2}%
\end{equation}
where $b=\frac{2\alpha}{\pi}(L-1).$ To the lowest order in $b$ and in the
region $x_{D}>>\nu^{2}=4m^{2}/M^{2},$ the above expression is in agreement
with the leading order expression \cite{B83}
\begin{equation}
\frac{1}{R_{0}}\frac{dR_{\mathrm{LO}}^{\mathrm{brem}}(x_{D})}{dx_{D}}%
=\frac{\alpha}{\pi}\frac{1}{1-x_{D}}\left\{  \left(  1+x_{D}^{2}\right)
\ln\left(  \frac{1+\beta_{x}}{1-\beta_{x}}\right)  -2x_{D}\beta_{x}\right\}
,\label{FLO}%
\end{equation}
where%
\[
\beta_{x}=\sqrt{1-\frac{\nu^{2}}{x_{D}}}.
\]
Thus we arrive at the differential rate
\begin{equation}
\frac{1}{R_{0}}\frac{dR_{\pi}^{RC}(x_{D})}{dx_{D}}=JF(x_{D}),\label{IB}%
\end{equation}
where the normalization factor $J$ takes into account to total RC. The
distribution (\ref{IB}), shown in Fig. 3, in contrast to (\ref{FLO}) is free
of nonintegrable singularity at $x_{D}=1$. So, we see the importance of taking
into account the higher orders of perturbation theory.

\begin{figure}[tbh]
\includegraphics[scale=1.5]{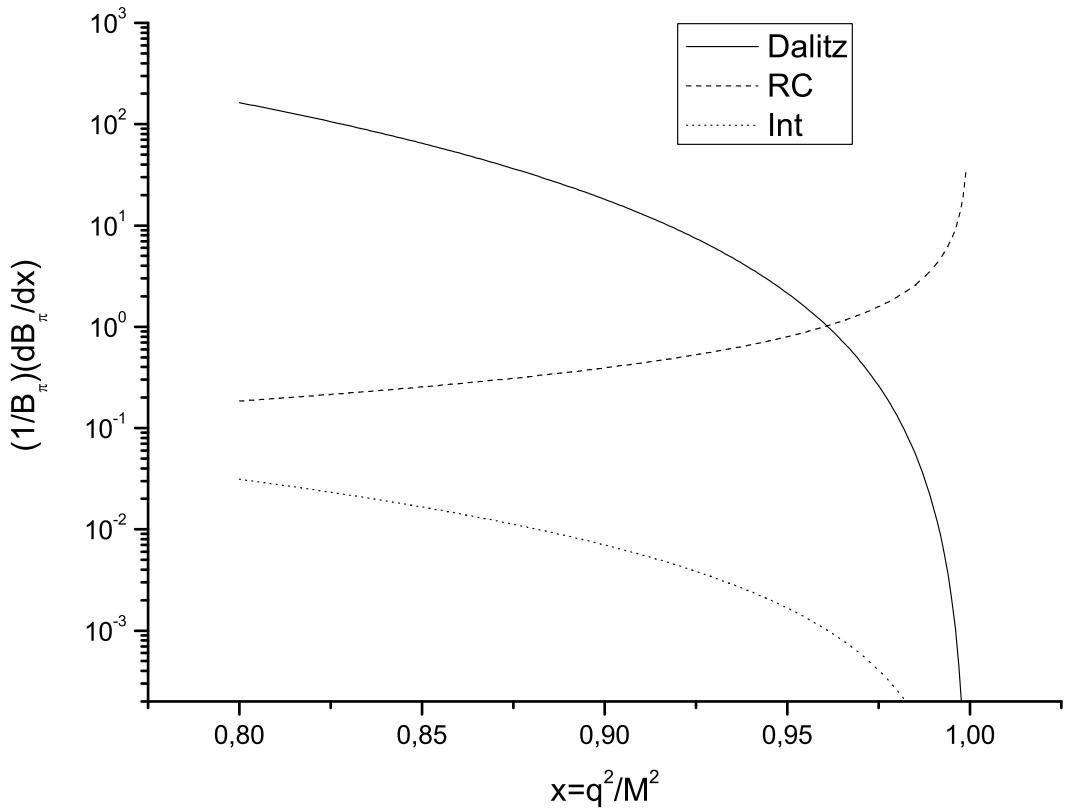}\caption{Distribution of different
contributions over $x=\frac{M_{ee}^{2}}{M^{2}}$: solid line - Dalitz mechanism
contribution, dashed line - inner bremsstrahlung contribution (\ref{IB}),
dotted line - interference of these two mechanisms.}%
\end{figure}

Adding the virtual and real photon emission contributions we finally obtain
\begin{align}
\frac{R_{\pi}^{RC}}{R_{0}} &  =J\int_{0}^{1}F(x_{D},L)dx_{D}=1+\alpha
_{virt}+\alpha_{soft}+\alpha_{hard}\nonumber\\
&  =1-\frac{\alpha}{\pi}\left[  \frac{1}{24}L^{2}+\frac{3}{4}L-\frac{\pi^{2}%
}{2}+\frac{21}{4}\right]  \approx0.968.\label{RadBS}%
\end{align}

The total radiative correction contains the large logarithm term $\sim L$
because the $\gamma_{5}$ current is not conserved. Compared with \cite{B83} we
provide a more detailed analysis of the effective $\pi^{0}\rightarrow
e^{+}e^{-}$ vertex revealing its DL structure\footnote{It is naturally to
expect the existence of DL-type RC in higher orders of perturbative theory. We
do not touch this problem here.}. However, numerically the ratio of the total
RC corrections to the lowest level rate estimated in (\ref{RadBS}) is very
close to $-3.4\%$ found in \cite{B83} and used by the KTeV Collaboration in
their analysis.

The branching ratio of the pion decay into an electron-positron pair has been
measured by the KTeV collaboration in the restricted kinematic region in order
to avoid a large background from the Dalitz process dominating at lower values
$x_{D}$. By using the distribution (\ref{F2}) we can estimate the factor of
extrapolation of the full radiative tail beyond $x_{D}>0.95$ as $f_{0.95}%
=1.114$. With this factor and scaling the result back up by the overall
radiative correction (\ref{RadBS}) we confirm the result (\ref{Bktev})
obtained by the KTeV Collaboration.

\section{Conclusions}

In this work, we reconsidered the contribution of QED radiative corrections to
the $\pi^{0}\rightarrow e^{+}e^{-}$ decay which must be taken into account
when comparing the theoretical prediction (\ref{Bth}) with experimental result
(\ref{BktevX}). Comparing with earlier calculations \cite{B83}, the main
progress is in detailed consideration of the $\gamma^{\ast}\gamma^{\ast
}\rightarrow e^{+}e^{-}$ subprocess and revealing of dynamics of large and
small distances. The large distance subprocess associated with final state
interaction produces the terms linear in the large logarithm parameter $L$.
The double logarithmic contributions ($\sim L^{2}$) correspond to
configurations when the particles in the loop are highly virtual. The total
result is in reduction of the normalization factor by $1-J\approx0.032$.
Occasionally, this number agrees well with the earlier prediction based on
calculations \cite{B83} and thus we confirm the KTeV analysis of RC factors.
So our main conclusion is that taking into account of radiative corrections is
unable to reduce the discrepancy between the theoretical prediction for the
decay rate (\ref{Bth}) and experimental result (\ref{Bktev}). Further
independent experimental and theoretical efforts are necessary. Note, that if
the discrepancy will stand as it is, then the possible explanation of the
effect is due the contribution to the decay width of low mass ($\sim10$ MeV)
vector boson appearing in some models of dark matter \cite{Kahn:2007ru}.

The question about the corrections is also important for the decays of $\eta$
and $K$ mesons to $\mu^{+}\mu^{-}$ and must be taken into account for the
analysis of experimental data. However, for these decays the large logarithm
parameter does not arise. The analysis of the lowest order RC to the decay
width of kaon to muon pair was given in \cite{GP74}. Unfortunately, the result
of \cite{GP74} explicitly depends on infrared singularities and cannot be used
in practice. Thus, in order to extract the information about $P\rightarrow
l\bar{l}$ decays (where $P=K_{L},\pi_{0},\eta,\dots$ and $l=e,\mu$) calculated
within the frame of definite models, the RC must be taken into account when
working with experimental data, e.g., for the process $\eta\rightarrow\mu
^{+}\mu^{-}$ \cite{egle}. For the process $\eta\rightarrow e^{+}e^{-}$ our
results are applicable and we estimate RC\ at the level $4.3\%$.

Finally, we have to emphasize that the role of RC is rather timely because of
the growing accuracy of modern experiments.

\section{Acknowledgments}

We are grateful to A.B. Arbuzov, M.A. Ivanov, N. I. Kochelev, S.V. Mikhailov
for helpful discussions on the subject of this work. The work (E.A.K., Yu.M.B.
and M. S.) is partially supported by the INTAS grant 05-1000008-8328, and
A.E.D. acknowledges partial support from the JINR-INFN program and the
Scientific School grant 4476.2006.2.

\end{document}